# From Strong To Weak Coupling Regime In A Single GaN Microwire Up To Room Temperature

A Trichet[1], F Médard[1], J Zúñiga-Pérez[2], B Alloing[2] and M Richard[1]
[1]Institut Néel, CNRS-CEA, BP 166, F-38042 Grenoble, France
[2]CRHEA-CNRS, Rue Bernard Grégory, F-06560 Valbonne, France

E-mail: Aurelien.trichet@grenoble.cnrs.fr

**Abstract**. Large bandgap semiconductor microwires constitute an advantageous alternative to planar microcavities in the context of room temperature strong coupling regime between excitons and light. In this work, we demonstrate that in the undoped half of a single GaN microwire, the strong coupling regime is achieved up to room temperature with a large Rabi splitting of 115 meV. The demonstration relies on a method which does not require any knowledge *a priori* on the energy of the uncoupled whispering gallery modes in the microwire, i.e. the details of the microwire cross-section shape. The other half of the microwire is heavily *n*-doped. Thus, within the same microwire, the exciton oscillator strength transits from its nominal value to zero within a region of 2 micron length. Using this property, we can observe the dispersion properties of a given whispering gallery mode both in the strong and weak coupling regime.

## 1. Introduction

In adequately designed semiconductor microstructures, the excitonic transition can be in the strong coupling regime with the electromagnetic field confined in a given spatial mode [1,2]. In this regime, the proper eigenstates of the system are exciton-polaritons, integer spin quasi-particles of mixed exciton and photon nature. With a very low mass and a bosonic nature, polaritons in planar microcavities have shown very interesting behavior at low temperatures like polariton lasing, Bose-Einstein condensation and superfluidity with original properties [3-6].

A large effort is currently devoted to obtain and study the strong coupling regime at room temperature. The main strategy consists in designing and fabricating microcavities using large bandgap materials like nitrides and zinc-oxide, where the exciton is stable at room temperature. One of the major outcomes would be the achievement of polariton lasing at room temperature under electrical pumping. With nitrides and zinc-oxide microcavities, strong coupling regime [7-9] and polariton lasing [10] at room temperature have been realized already. However, electrical injection within microcavities is a challenging task: free carriers must reach the quantum wells placed at the heart of the structure, i.e. in the cavity layer. This requires the fabrication of *p* and *n* doped Bragg mirrors or advanced etching techniques for intracavity electronic injection [11]. In addition, III-N and ZnO-based microcavities exhibit lower Rabi splitting than microwires due to the significant fraction of the electromagnetic field contained in the Bragg mirrors, and by the presence of internal electric fields at every heterointerface (polarization discontinuities of spontaneous and piezoelectric nature) that can reduce the exciton oscillator strength by quantum confined stark effect [12].

Large bandgap microwires constitute a very interesting alternative approach to planar microcavities. Recently, ZnO microwires have been shown to sustain high quality 1-dimensional polariton gas at room temperature with giant Rabi splitting and narrow linewidth [13]. In these structures, the photon is efficiently confined within the microwire by total internal reflection at the six semiconductor/air interfaces of an hexagonal cross section (this hexagonal shape arises due to the crystalline wurtzite structure). These so-called "hexagonal whispering gallery modes" (HWGMs) with high-quality factors up to 800 have close to 100% overlap $\alpha$ with the free excitons which can result in strong coupling regime and giant Rabi splitting. However, in the context of polariton lasing, ZnO suffers from the well-known issue of *p*-doping [14,15] which prevents any electrical injection as long as it is not solved.

In this context, reaching the strong coupling regime in GaN microwires is of high interest for two main reasons: first, the growth of this material is technologically more mature, thus more complex structures with embedded quantum wells or quantum-dots in the strongly coupled microwire can be envisaged on a short-term. Second, unlike ZnO, nitrides-based microwires can be nowadays designed into p-i-n diodes for electrical injection [16]. In this report, we show that GaN microwires grown by metal-organic vapor phase epitaxy (MOVPE) are in the exciton-photon strong coupling regime from cryogenic up to room temperature, with a giant Rabi splitting of 115 meV and a linewidth typically 15 times smaller. This Rabi splitting exceeds state of the art nitride-based planar microcavities [7,10]. Interestingly, these microwires give us the opportunity to experimentally switch off the strong coupling regime by removing the exciton levels. This is done by comparing within the same single wire an undoped segment and an intentionally highly *n*-doped segment.

GaN microwires have been grown by MOVPE on *c*-plane sapphire substrates. In order to promote the vertical growth along <0001>, and after a nucleation step in which truncated GaN pyramids are formed, silane is injected into the reactor during the first 30 min of the growth together with $NH_3$ and TMGa [17]. As a result, the first half of the microwire is highly *n*-doped. Then, silane injection is stopped so that the upper part of the wire is nominally undoped. During this step, the TMGa to $NH_3$ ratio (i.e. the V/III ratio) is much lower than that conventionally used for two-dimensional growth. This growth procedure results in a transition region, 2 to 3 micrometer thick, at the interface between the doped and undoped segments of the wire (see Fig.1.b $z$=52 μm to $z$=55 μm). The microwires thus obtained feature 2 to 10 micrometers in diameter and 20 to 40 micrometers in length. In all cases, the wurtzite GaN c-axis coincides with the wire longitudinal axis (referred to z in this article).

We performed optical micro-photoluminescence measurements on single GaN microwires with spatial and angular resolution. To do so, we harvested the wires from their substrate and dispersed them onto a glass plate, which was then placed in a variable temperature optistat. Optical excitation of single wires was realized with a CW He-Cd laser at 325 nm, i.e. always above the band to band transition of bulk GaN at 5 K and 300 K. Broadband emission is thus obtained below the bandgap due to lower polariton branches, LO-Phonon replica, optically active impurities and surface states. The latter largely contribute to the background emission observed on top of polariton emission in Fig.1.c. [18]. For the micro-photoluminescence experiment shown in Fig.1, the laser was exciting the whole microwire, and the photoluminescence image in real space was formed on the entrance slit of a 1-meter focal length monochromator. For angle resolved measurements, the laser was focused into a 2-micron spot by the near-UV refractive objective so that a short segment of the wire featuring constant diameter (e.g. $z$=47 μm and $z$= 65μm on the wire shown Fig.1) is selectively excited. The photoluminescence image in reciprocal space is formed on the slit of the monochromator, with emission angle $\theta$ (Fig.1) varying from -30° to +30° along the slit of the monochromator (objective NA=0.5). The photoluminescence passes through a linear polarizer in order to select σ-polarized (electric field perpendicular to c-axis) emission.

Let us focus first on the micro-photoluminescence spectra shown in Fig.1.b and Fig.1.c which were obtained at T=10 K. The most striking feature is the ensemble of spectrally narrow bent stripes separated by a few tens of meV from each other and visible over the whole length of the wire. In the upper part of the wire (55 µm<z<71 µm) which is nominally undoped, they correspond to HWGMs strongly coupled to free excitons (the peaks on Fig.1.c grey line) as will be discussed below. A weak contribution of A and B exciton (3477 meV and 3482 meV) can also be seen in this region. In the lower part of the wire (43 µm<z<52 µm) the wire is heavily *n*-doped, with a ~1-2.10$^{20}$ cm$^{-3}$ rough estimate based on the high energy edge (at 3555 meV) of the photoluminescence [19], which corresponds roughly to an electron Fermi energy 60 meV above the bottom of the conduction band in GaN. In this region, we show that the HWGMs are in the weak coupling regime.

## 2. Spectroscopy of an undoped segment of a single microwire

In order to demonstrate the strong coupling regime, we measured the angular dispersion (i.e. the energy *E* versus the emission angle *θ*) of the emission of several HWGMs in the undoped segment of a single microwire. We chose to work with large wire diameters (3-7 microns) for two reasons: first, the larger the diameter, the higher the quality factor of the HWGM at a given energy [20]; second, the low free spectral range $\Delta=E_{m+1}-E_m$ between two consecutive HWGMs, of 23.5 meV, provides us with about 20 well separated HWGMs within a detection spectral window of [3150,3550] meV. The latter is chosen in order to include free excitonic transitions on the high energy side (A, B and C excitons are in the range 3477 meV to 3499 meV at 10 K), several lower polariton branches in the middle energy range, and quasi-pure (i.e. with negligible coupling with the excitons due to exciton-photon detuning much larger than the Rabi splitting) HWGMs on the low energy side.

The results are shown on the upper panels of Fig.2 and Fig.3 for temperatures of 10 K and 300 K respectively. In this measurement, only σ-polarized emission is detected (with electric field perpendicular to c-axis). Another family of polaritons is observed for π-polarization [13,21], with a larger contribution of the C exciton to the total oscillator strength, a lower one from A and B, and a similar overall Rabi splitting [22]. But due to poor thermal population of the C exciton at cryogenic temperature, we found that the π-polariton polariton luminescence was too dim to be exploited quantitatively (not shown).

σ-polarized angle-resolved luminescence shows several well-resolved dispersion branches with decreasing curvature from low (Fig2.c, bottom) to high (Fig.2.a, top) emission energy. We show below that a quantitative measurement of this curvature change versus energy provides an unambiguous demonstration of the strong coupling regime between exciton levels and HWGMs in the microwire. To confirm this interpretation, in section III, we check experimentally that upon switching off the excitonic oscillator strength using the doped region of the microwire, the strong coupling is also switched off.

Another interesting feature is the sharp linewidth of the modes at the lowest measured energy (around 3.2 eV) where it is of nearly pure photonic origin. It corresponds to a quality factor of HWGM of 600-700. Note that this value is a lower bound only since the measured linewidth is not always homogeneous. Indeed, on top of (and sometime superimposed to) the brightest HWGMs, dimmer ones are visible (mostly in Fig.3) with different energy spacing, which result possibly from a non-perfectly hexagonal cross-section. Note also that in hexagonal microwires the quality factor varies strongly from mode to mode due to the details of the field distribution with respect to the six corners. Moreover the behavior of the HWGMs linewidth versus angle is also non-trivial: in cylindrical microcavities a strong TE/TM mixing occurs at non-zero angle which is reinforced by the excitonic anisotropy and the strong coupling regime [21]. Furthermore, at higher energy and due to the excitonic fraction, the linewidth includes a thermal component due to interaction with phonons. For example, for polaritons with ~50% excitonic fraction: a linewidth of 6.5 meV is found at 5 K which increases to 7.5 meV at 300 K (mode above the green dotted lines in Fig.2 and Fig.3). This small thermal contribution to the linewidth is due to the partial decoupling of these modes from the thermal bath of phonons. Like in [13], a complete polariton scattering calculation including the acoustic and optical phonon population would be required to understand it quantitatively. Qualitatively, this narrow linewidth at room temperature is consistent with the suppressed scattering with acoustic phonons. In any case, we do not enter into the details of these mechanisms since the linewidth dependence on angle or energy does not affect our measurements or our conclusions on the regime of light-matter interaction.

*2.1. Method to determine the exciton-photon coupling regime and the Rabi splitting in a single microwire of large diameter based on dispersion curvature measurement.*

Unlike our previous work carried out at room temperature [13], due to too short space between the cryostat and the objective, we cannot access large emission angle, and observe the inflexion point in the lower polariton dispersion which is a characteristic signature of the strong coupling regime. However, it is important to stress that this inflexion point is only the consequence of the resonance in the dielectric function that exists in the strong coupling regime at the excitonic energy, as described by Eq.(2). Other equally significant and observable consequences of this resonance are the anti-crossing between the upper and lower polaritons (not observable here due to thermal relaxation of upper polariton), and the lower polariton dispersion flattening for increasing energy (i.e. divergence of the lower polariton mass upon approaching the exciton energy). It is this latter signature that we exploit here quantitatively to characterize the strong coupling regime.

In a microwire of a few microns in diameter, HWGMs are confined within the cross-section so that in this plane, the wavevector is quantized, leading to discrete modes of the electromagnetic field (labeled by a single value *m* throughout this paper for simplification). On the other hand, due to translational invariance along the wire axis *z*, the wave vector $k_z$ is a good quantum number describing free motion of the field in this direction [21]. Owing to the conservation of energy and $k_z$, the energy of the *m*-th HWGMs reads:

$$E_m(\theta) = E_{0,m}\left(1 - \frac{\sin^2\theta}{\varepsilon(E_m)}\right)^{-\frac{1}{2}}, \quad (1)$$

where $E_{0,m}$ is the energy of the *m*-th HWGM at $\theta = 0°$ and $\varepsilon(E_m)$ is the dielectric function for σ-polarized light. Thus, each HWGM, provides in principle a tight regular sampling of the dielectric function in the detection window. When the excitonic response is disregarded, $\varepsilon(E_m) = \varepsilon_\infty$ the background dielectric constant which accounts for every other contribution, and $E_m(\theta) = \tilde{E}_m(\theta)$ the dispersion of pure HWGMs, decoupled from the excitons. In the framework of the linear response theory, the electromagnetic response of the excitonic transitions can be explicitly added to the dielectric function $\varepsilon(E)$, under the form [23]:

$$\varepsilon(E) = \varepsilon_\infty\left(1 + \sum_{j=A,B,C} \frac{\alpha^2 \Omega_j^2}{E_j^2 - E^2 + i\Gamma_j E}\right), \quad (2)$$

where $\Omega_j$ is the contribution to the Rabi splitting of the excitonic transition *j=A, B* and *C* of energy $E_j$. $\alpha$ is the spatial overlap integral between the excitonic medium and the HWGMs, and $\Gamma_j$ is the non-radiative linewidth of the exciton levels. $\alpha^2$ is lower than 1 due to the evanescent part of the HWGMs and to an excitonic dead layer located right below the air-semiconductor interfaces [23,24].

In our measurement spectral window, the detected energy $E$ always remains close to $E_j$ and $\Gamma_j \ll 2(E - E_j)$. In these conditions $\Gamma_j$ does not affect the shape of the polariton dispersion, only its linewidth, and will be set to 0 in our case.

Using a single set of parameters for each investigated temperatures, we can fit properly the entire set of dispersion branches of Fig.2 and Fig.3 (upper panels). Due to the maximum detection angle of 30° of our apparatus, the measured dispersion branches hardly deviate from a parabolic shape. Thus, the fitting procedure amounts to determining quantitatively the curvature of each dispersion branch.

We take the exciton energies and oscillator strengths from the literature (displayed in Table 1), assuming unstrained GaN. Indeed, TEM measurements indicate that the strains are mostly relaxed in the first 200nm of the wire. Note that the presence of residual strain would not affect our conclusions anyway: strain typically shifts the exciton energies by 3-4 meV [25] and changes by a few tens of percent the oscillator strength ratio (total oscillator strength remains unchanged) between B and C excitons [26]. Considering the 18meV splitting between B and C excitons, these energy scales remain negligible as compared to the Rabi splitting and can be disregarded.

Thus, only two parameters are left free in the fitting procedure: $\alpha$ and $\varepsilon_\infty$. These two parameters can be fitted independently from each other since for polariton modes far below the exciton energies as compared to the Rabi splittings $\Omega'_j$, the right term of Eq.(2) vanishes and the dispersion branch is sensitive only to $\varepsilon_\infty$. It is the case of the modes at the bottom of Fig.2.c and Fig.3.c.

Note that this procedure has a significant advantage over other methods since it does not require any knowledge *a priori* on the energy of the uncoupled confined electromagnetic modes $\tilde{E}_{0,m}$ versus mode number *m* in the microwire, which depends a lot on the precise shape of its cross-section. Instead, in our method the spectrum $\tilde{E}_{0,m}$ is a result of the procedure since it is connected with the measured energy of the polariton modes $E_{0,m}$ by the relation:

$$E_{0,m} = \frac{n_\infty}{n(E_{0,m})} \tilde{E}_{0,m} \qquad (3)$$

*2.2. Results and discussion on the exciton-photon coupling regime in a single microwire*

We applied this method to the measurements shown in Fig.2 and Fig.3 (upper panels, for the undoped region). As explained above the already known parameters of the fitting procedure are the exciton energy and oscillator strength (table 1), plus the set of energies $E_{0,m}$ for each dispersion branch *m* which are provided by the measurement. Then the whole set of dispersion branches are fitted with the calculated ones leaving only two parameters free: $\alpha$ and $n_\infty$. To be more specific, in a first step, the background index $n_\infty$ alone is determined by fitting the lowest energy HWGMs with the calculated ones. In this energy range, $\alpha$ has no effect on the calculated dispersions; therefore this is a one-parameter fit. The uncertainty associated to $n_\infty$ is defined as the one resulting in a deviation of ± half a linewidth of the fit with respect to the measured curves at θ=±30°. In a second step, the same strategy is applied for $\alpha$, $n_\infty$ being now fixed. One given value of $\alpha$ provides a good fit for all branches of the data set. The uncertainty of $\alpha$ is lowered if narrow polariton branches of energy close to the excitons can be used for the fit. At room temperature, the polariton branches normally situated in the range [3340, 3410] meV are not visible (Fig.2.d). The reason is that at room temperature the emission of these modes is quenched and broadened due to the scattering of polaritons with the thermal bath of phonons toward the free exciton states [13]. This is the reason why the uncertainty is large for the extracted $\alpha$ at room temperature. Note that since the linewidth of a polariton mode at a given angle is always larger than the standard deviation of its mean energy this method provides rather an overestimate of the uncertainties.

The thus extracted values and the total Rabi splitting are summarized in Table 1: a large background index $n_\infty$=3.1 and an overlap integral $\alpha$=92% are found both at T=10 K and at room temperature. This background index is larger than 2.6 usually found in the literature in the near-UV [25,27]. Likely contribution could be the significant overlap of the HWGM with a large density of deep electronic levels (contained within the bandgap) located at the surface of the wire [29]. Note that according to Eq. (2) the value of the background index is a factor which rescales the excitonic contribution to the dielectric function. Therefore its value does not influence the extraction of the Rabi splitting.

**Table 1.** Exciton energies ($E_j$) and contributions to the Rabi splitting ($\Omega_j$) taken from the literature assuming unstrained GaN. In italic is shown the result of the fitting procedure: $\alpha$ is the normalized spatial overlap integral between exciton and HWGM and $n_\infty$ is the background index. The resulting overall Rabi splitting $\Omega$ is shown in bold characters.

| T (K) | $E_A$ (meV) | $\Omega_A$ (meV) | $E_B$ (meV) | $\Omega_B$ (meV) | $E_C$ (meV) | $\Omega_C$ (meV) | $n_\infty$ | $\alpha$ |
|---|---|---|---|---|---|---|---|---|
| 5K | 3477 [29] | 90 [23] | 3482 [29] | 78 [23] | 3499 [29] | 37 [23] | *3.1±0.1* | *0.92±0.1* |
| 300K | 3409 [30] | 90 [23] | 3414 [30] | 78 [23] | 3432 [30] | 37 [23] | *3.1±0.1* | *0.92±0.3* |

| T (K) | $\Omega$ (meV) | $\Gamma_0$ (meV) |
|---|---|---|
| 5K | **115±10** | 6.5±1 |
| 300K | **115±40** | 7.5±1 |

The large value of $\alpha$ is the first main result of this article since it clearly shows that the strong coupling regime is achieved at 10 K and 300 K. The unambiguous demonstration requires the comparison between the overall Rabi splitting $\Omega$ and the linewidth $\Gamma_0$ of the polariton mode around the resonance (i.e. the region of detuning where the anti-crossing takes place between the exciton and the photon energies) [23]. $\Omega = \alpha\sqrt{\Omega_A^2 + \Omega_B^2 + \Omega_C^2}$ is a proper definition of the overall Rabi splitting when the latter is much larger than the energy separation between the different exciton levels. We find $\Omega = 115\pm10$ meV at 10 K, and $\Omega = 115\pm40$ meV at room temperature, and both meet this condition. The linewidth $\Gamma_0$ of the mode at $\delta\sim0$ (cf. Fig.2.a at T=10 K and Fig.3.a at room temperature) is fitted with a Lorentz lineshape by carefully removing the background emission, for which we have taken advantage of the fact that this emission is not angle dependent. We find $\Gamma_0$=6.5±1 meV at T=10 K and $\Gamma_0$=7.5±1 meV at room temperature. Finally, comparing $\Gamma_0$ and $\Omega$ at cryogenic and room temperatures shows unambiguously that the strong coupling regime is achieved in both cases with a figure of merit (defined as the ratio between Rabi splitting and polariton linewidth) of 20 at 10 K and 15 at room temperature.

Like in [13] the very fact that each polariton branch *m* is well separated from the neighboring ones (*m*±1) in energy indicates that every branch is associated to a single transverse mode with $k_z$ as only degree of freedom. As a result, polaritons in our GaN microwires also have a one-dimensional character.

An alternative way to look at these results is shown in Fig.4.a (T=10 K) and Fig.4.b (room temperature), where the polariton modes energy $E_{0,m}$ is plotted versus the mode number *m*. The uncoupled modes energies $\tilde{E}_{0,m}$ extracted using Eq.(3) in the procedure discussed above are also shown. Interestingly, the latter appear equally spaced in energy by $\Delta$=23.5±0.3 meV (cf. Fig.4.a and Fig.4.b, hollow circles plots). Indeed, in the limit of modes of large angular momentum *m*, $\Delta$ is then simply connected with the radius of the microwire and the background index by the relation $\Delta = hc/6n_\infty R$ [20], where R is the small radius of the hexagon. R=2.7 µm is found in agreement with optical microscope measurements (not shown). As expected from the strong coupling regime, the lower polariton energies $E_{0,m}$ plotted versus *m* exhibit an anti-crossing behavior with the free exciton energies. This is not the case in the weak coupling regime shown in the same plot (Fig4.a and Fig.4.b hollow squares) and discussed in the next section.

Using Eq.(3), one can also plot the normalized dielectric functions $\varepsilon(E)/\varepsilon_\infty$. This is shown in Fig.4.c and Fig.4.d (filled circles). The shape of this normalized dielectric function depends only on the regime of coupling with the excitons: in the strong coupling regime we observe as expected a strong divergence upon approaching the exciton energy as described by Eq.(2) for vanishing Γ. On the same plot is shown the dielectric function in the weak coupling regime (Fig.4.c and Fig.4.d, hollow squares). As discussed in the next section, in this situation it is only weakly perturbed.

## 3. Spectroscopy of the heavily n-doped segment of a single microwire

Heavy n-doping is an interesting situation in the context of the strong coupling regime between exciton and photon because of its effect on the excitons: upon increasing the free charges density, the exciton oscillator strength strongly decreases and the exciton emission gets spectrally broader. At large doping level, like in the present case, bound electron-hole states do not exist anymore. Instead, correlated many-electrons plus one-hole states are present (sometime called "Mahan exciton" or "Fermi edge singularity") which exhibit a strongly reduced oscillator strength as compared to the exciton one in the undoped case, but still slightly enhanced with respect to uncorrelated electron-hole recombination in the undoped case [31,32]. Far beyond the Mott transition, these states have a recombination energy largely blueshifted with respect to the undoped band to band transition. Indeed in this regime, the bandgap renormalization (redshift due to Coulomb interaction between free electrons) is dominated by the conduction band filling.

In our case, in the *n*-doped region of the wire, we observe a high energy edge of photoluminescence at 3555meV which correspond to an electron Fermi energy 60 mev above the bottom of the conduction band in GaN (see Fig.1, right panel). This behavior is consistent with the aforementioned regime of very large doping density. Indeed the high energy photoluminescence (between 3480meV and 3560meV) results from the recombination between a photogenerated hole and any electron between the conduction band edge and the Fermi energy. This mechanism involves non *k*-conserving recombinations which are allowed in the degenerate regime due to break-up of the single electron excitation picture [33]. The important point in the context of this work is that in this regime, the excitonic levels with the energy and oscillator strength characteristic of undoped GaN are completely suppressed, i.e. the strong coupling regime between excitons and HWGMs is switched off.

We performed angle-resolved spectroscopy of the same microwire as in the previous section, but this time on the heavily n-doped segment to measure how the strong coupling regime is affected by the suppression of the excitonic transitions. The results are shown on the lower panels of Fig.2 and Fig.3. Well-resolved dispersion branches of pure HWGMs are observed. Their linewidth is as narrow as that of the undoped segment, i.e. in the 5 meV range. This is true even for modes lying above the energy of the undoped GaN exciton. Indeed, no absorption induced broadening is expected [32] in heavily n-doped GaN.

Regarding the shape of the dispersion branches, their curvature in the vicinity of the band edge is much less flattened than in the undoped region. Using Eq.(1), we could determine the dielectric function $\varepsilon(E)$ by fitting each dispersion branch independently (blue solid lines in Fig.2 and Fig.3 lower panels). As shown in Fig.4, all the features of the weak coupling regime are recovered both at T=10K and at room temperature: no anti-crossing with the exciton levels is observed in the plot of the HWGMs energy versus *m* (Fig.4.a and Fig4.b hollow squares), and the dielectric function exhibits no divergence either at the exciton energies (Fig.4.c and Fig4.d hollow squares). A slowly increasing $\varepsilon(E)$ is observed instead, likely due to onset of absorption at the energy of the transition from the valence band to the Fermi level in the renormalized conduction band [32].

Note that in *n*-doped (electron densities up to $10^{19}$ cm$^{-3}$) bulk GaN a broad resonance with slightly enhanced oscillator strength (with respect to the band-to-band transition) has been already reported and attributed to Fermi edge singularity [32]. The latter could possibly participate in the gentle increase of the dielectric function that we observe at high energy.

The direct comparison of both doped and undoped region of the same microwire is the second main result of this article. Everything else being identical, one can observe quantitatively how removing the excitonic transitions affect the dielectric function and results in switching off the strong coupling regime with the HWGMs.

As a concluding remark, let us mention that we attempted to reach the regime of polariton lasing in the undoped part [34] of the microwire as well as the regular lasing regime in the doped part both at T=10K and at room temperature. 100-femtoseconds pulses at 351nm have been used with a maximum average power of 75mW focused on 2 micron diameter spot. However, the quality of GaN in our microwires was not good enough to achieve efficient relaxation of the reservoir excitons towards the polariton branches. Instead, time-resolved measurement of the polariton PL versus excitation power (not shown) reveals that the excitons relax much faster towards some unidentified non-radiative impurity-bound states or surface states where they are lost for polariton excitations. This mechanism prevents reaching the threshold as long as these defect states are not saturated. We are convinced that this problem will be overcome soon by improving the material quality.

## 4. Conclusion

In this work, using the energy dependent curvature of the polariton branches we have shown that MOVPE-grown GaN microwires are in the 1-dimensional strong coupling regime from cryogenic up to room temperature with a Rabi splitting of 115 meV. We could experimentally verify this interpretation by removing within the same wire the excitonic levels by heavy electron doping. The strong coupling regime is then found to be completely switched off. The large Rabi splitting found in undoped GaN microwires opens interesting perspectives for the fabrication of polariton laser diodes [35] at room temperature, a device which operation is still to be demonstrated. Indeed, large bandgap materials like GaN are highly desirable for this application owing to the larger binding energy of the exciton. Furthermore it has been shown already that GaN microwires can be grown into a *p*-i-*n* junction [16], a device much simpler to fabricate than nitride-based planar microcavities designed for electronic injection.

**Acknowledgements**

MR and AT acknowledge enlightening discussions with L. S. Dang. MR and FM acknowledge support from the ERC (grant No. 258608 "Handy-Q"). AT acknowledges support of the French Nanoscience foundation (Project No. FCSN-2008-10JE "RICOPHIN"). JZP acknowledges support from the ERC (CLERMONT 4 grant No. FP7-PEOPLE-ITN-2008 235114).

**FIGURES**

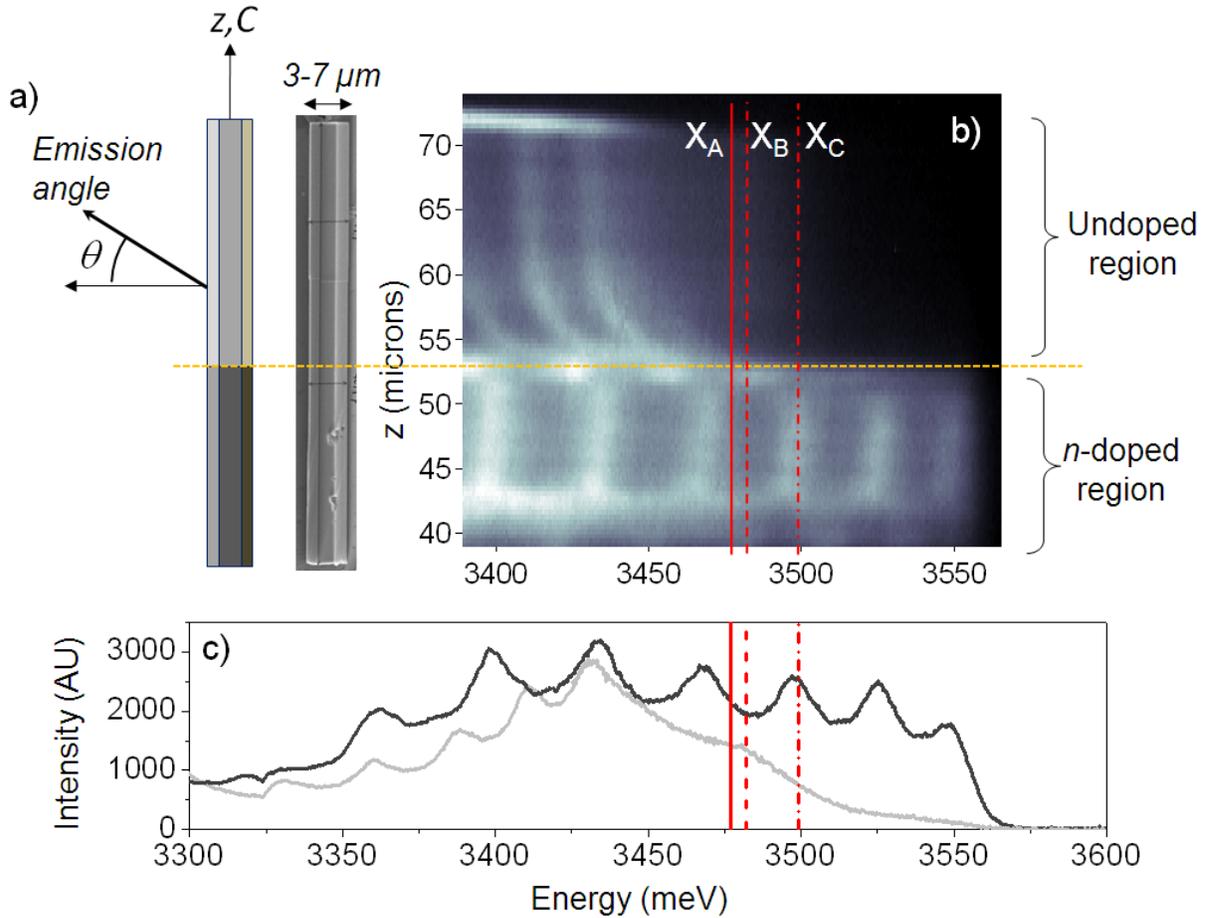

**Figure 1.** a) Schematic representation of a GaN microwire. The dark grey region represents the n-doped segment and the light grey one represents the undoped segment. Next to it is shown a typical SEM micrograph of a GaN microwire. b) micro-photoluminescence spectrum (photon energy on the horizontal axis, PL intensity is color scaled) versus position on the microwire (vertical axis z) in the near band edge region of GaN at T=10K. The bright bent lines are whispering gallery modes in the strong (undoped region, i.e. z=72µm to z=55µm) or weak coupling regime (doped region, i.e. z=52µm to z=43µm) depending on the position z. The horizontal orange dashed line show the position separating the doped from the undoped segments of the wire. The exciton levels of undoped GaN at T=10K are indicated by the red vertical lines. c) Emission spectra obtained from slices of measurement b) taken at z=65µm (light gray solid line, undoped segment) and z=47µm (black solid line, doped segment).

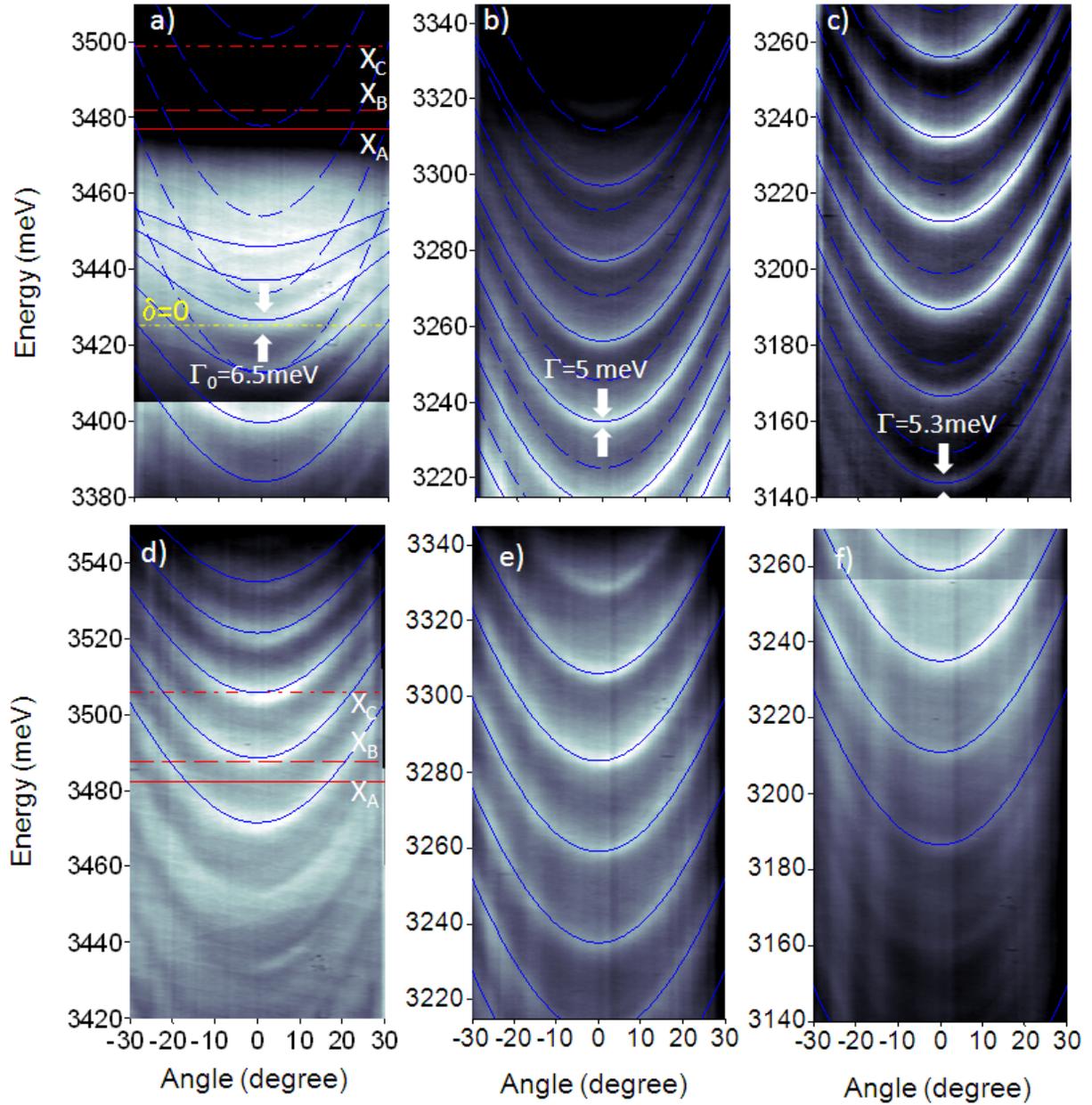

**Figure 2.** Measured σ-polarized photoluminescence intensity (color scale) of the undoped (upper panels a, b and c) and heavily n-doped (lower panels d, e and f) segment of the wire at T=10K with angular (x-axis) and spectral (y-axis) resolution. The free exciton energies labeled $X_A$, $X_B$ and $X_C$ are indicated as straight red lines (solid, dashed and dot-dashed respectively). The blue solid lines are the calculated dispersion branches in the strong coupling regime. The calculated uncoupled modes are shown in dashed lines. The contrast of the color scale has been sometime adjusted within the same image in order to maximize the visibility of the dispersion branches. In a) the energy corresponding to polaritons with zero detuning (δ=0) between excitons and uncoupled HWGM is indicated by the yellow dashed line. The linewidth Γ of selected modes is indicated in panel b) and c)

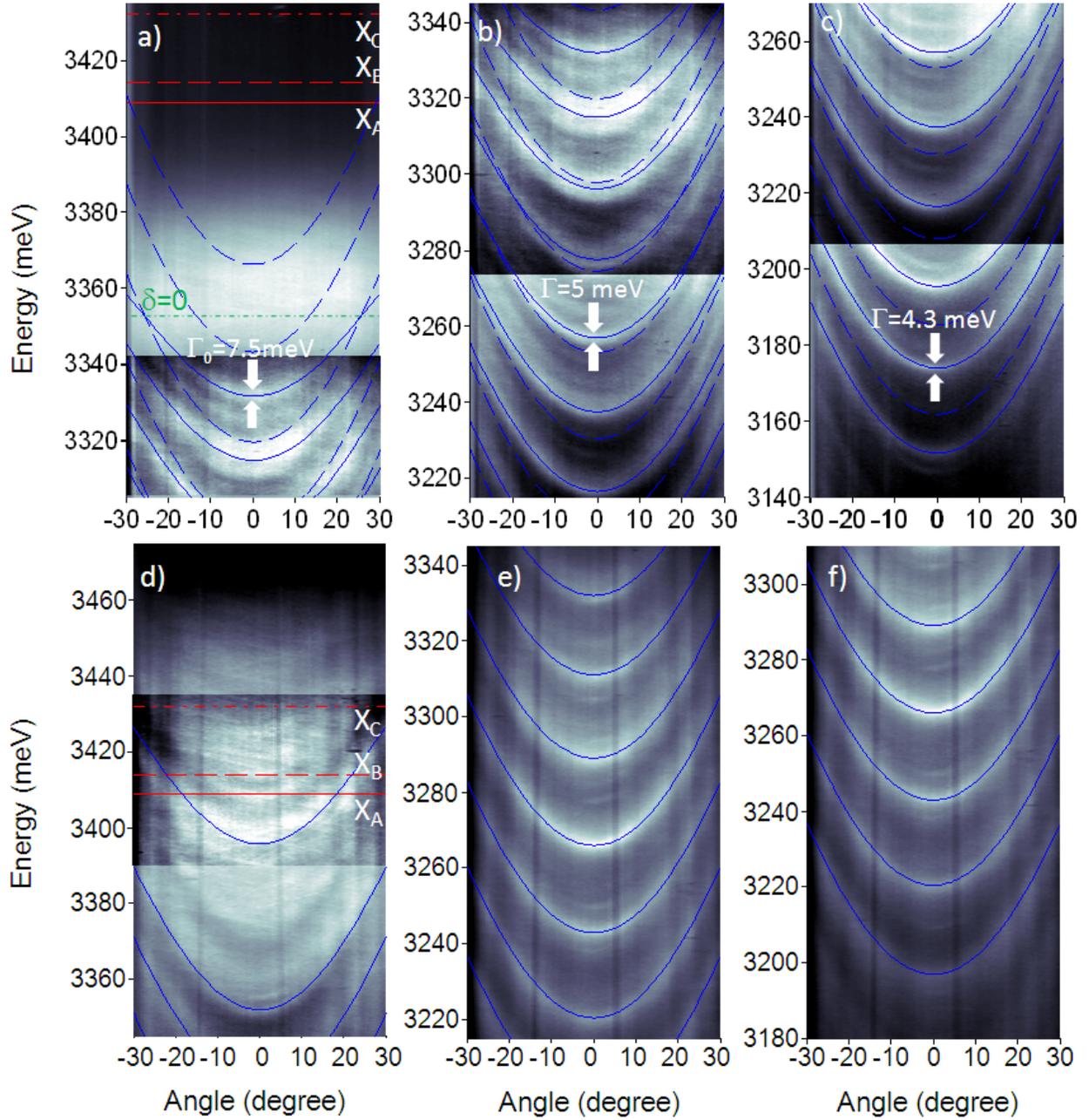

**Figure 3.** Like Fig.2 but at room temperature. The measurements have been done on the same microwire as that of Fig.2.

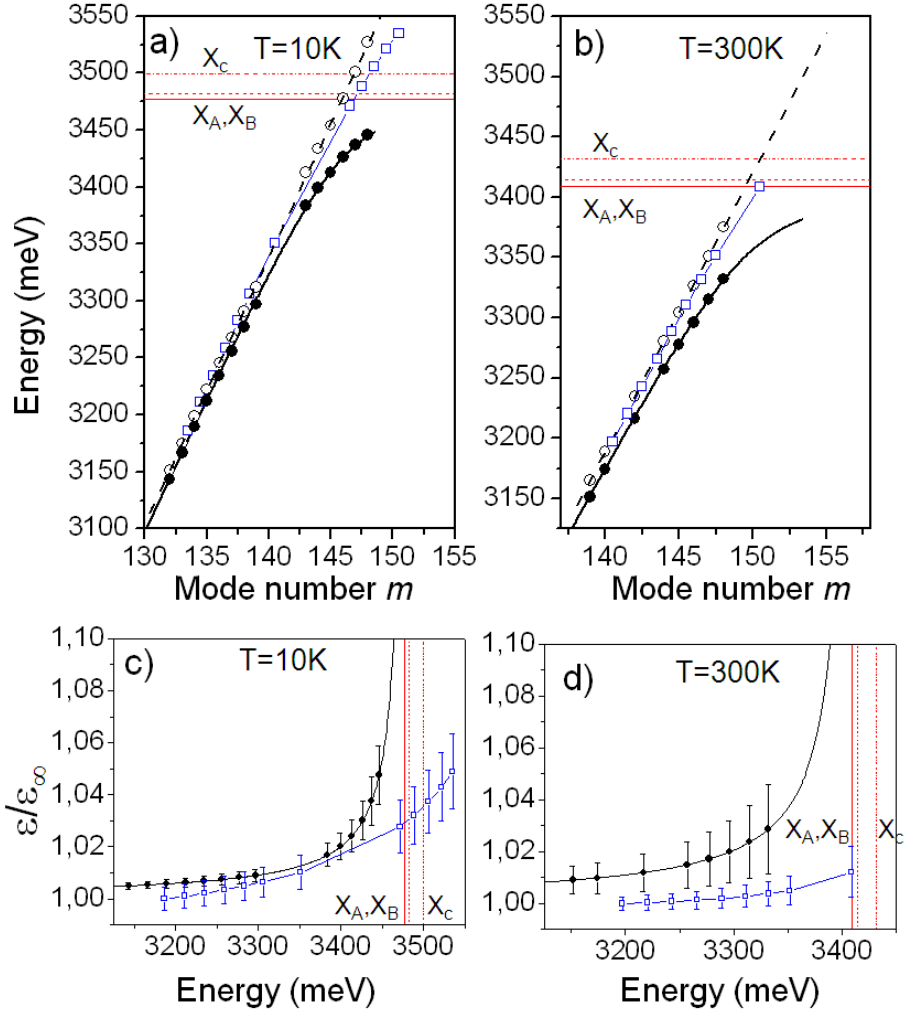

**Figure 4.** Measured polariton energy at θ=0 (black filled circles) at T=10K (a) and room temperature (b) versus mode number in the undoped segment. "Mode numbers" are integer numbers that simply label the modes, starting from 0 at zero energy. The black hollow circles are the calculated uncoupled HWGMs energy in the undoped segment (see text). Blue hollow squares are the HWGMs energy at θ=0 obtained in the doped segment. The black dashed line is a linear fit of the uncoupled HWGMs energy which provides a free spectral range of Δ=23.5±0.3meV (see text). The black solid line is the calculated polariton energies at θ=0, assuming uncoupled HWGMs equally spaced in energy by Δ. The exciton levels A, B and C are indicated. Lower panels show the measured and fitted (using Eq.(2) and the parameters listed in table (1)) values of the dielectric function in the undoped (black filled circles and black line respectively) at 10K (c) and room temperature (d). Hollow blue squares are values of the dielectric function obtained in the doped region (the first data points have been arbitrarily normalized to one). As explained in the text, the error bars are determined in a way that rather overstimates the uncertainty.

14